\documentclass{article}

\begin{document}

\title{\bf \Large Momentum Imparted by Gravitational Waves}

\author{M. Sharif \thanks{Present Address: Department of Mathematical Sciences,
University of Aberdeen, Kings College, Aberdeen AB24 3UE Scotland,
UK. $<$msharif@maths.abdn.ac.uk$>$}
\\ Department of Mathematics, University of the Punjab,\\ Quaid-e-Azam Campus
Lahore-54590, PAKISTAN.}

\date{}

\maketitle

\begin{abstract}
We calculate momentum imparted by colliding gravitational waves in
a closed Friedmann Robertson-Walker background and also by
gravitational waves with toroidal wavefronts using an operational
procedure. The results obtained for toroidal wavefronts are well
behaved and reduce to the spherical wavefronts for a special
choice.
\end{abstract}
{\bf PACS: 04.20.Cv, 04.30.-w}

\newpage

\section{Introduction}

The localization of energy and momentum has been one of the most
interesting and important problems in Einstein's theory of General
Relativity (GR). The different attempts at constructing an
energy-momentum density do not provide a generally accepted
definition. Consequently, different people have different points
of view. In one of his papers Cooperstock [1] argued that in GR,
energy and momentum are localized in regions of the non-vanishing
energy and momentum tensor. As a result gravitational waves are
not carriers of energy and momentum in vacuum. Since the
gravitational waves, by definition, have zero stress-energy
tensor. Thus the natural question about the existence of these
waves was raised. However, the theory of GR indicates the
existence of gravitational waves as solutions of Einstein's field
equations [2]. This contrary result arises because energy is not
well defined in GR.

Ehlers and Kundt [3], Pirani [4] and Weber and Wheeler [5]
resolved this problem for gravitational waves by considering a
sphere of test particles in the path of the waves. They showed
that when gravitational waves passed through the test particles
these acquired a constant momentum from the waves. However, the
procedure [6] does not provide a simple prescription that can be
used for arbitrary case. Qadir and Sharif [7] used an operational
procedure, embodying the same principle and found a closed form
formula which can be applied to arbitrary spacetimes. This
procedure provided the similar results when we applied to plane
and cylindrical gravitational waves [7]. In a recent work [8],
this approach has been used to find the momentum imparted by
gravitational waves with spherical wavefronts. Interestingly, we
obtain physically acceptable results which coincide with results
obtained by using M$\ddot{o}$ller's prescription. This paper
focuses the problem of finding the momentum imparted by colliding
gravitational waves in a closed Friedmann Robertson-Walker (FRW)
background and also by gravitational waves with toroidal
wavefronts.

We use an operational procedure for expressing the consequences of
Relativity in terms of the Newtonian concept of gravitational
force. The pseudo-Newtonian {\it gravitational force} is defined
as the vector whose intrinsic derivative along the separation
vector is the maximum tidal force, which is given by the
acceleration vector for preferred class of observers. We will not
discuss this formalism here as it is available in detail elsewhere
[9]. We shall restrict ourselves to briefly mention the essential
points of the formalism. In the free fall rest-frame the extended
pseudo-Newtonian four-vector force can be written in the form
[7,9]
\begin{equation}
F_0=m[\{\ln(A/\sqrt{g_{00}})\}_{,0}-g_{ij,0}g^{ij}_{,0}/4A],
~~F_i=m(\ln\sqrt{g_{00}})_{,i},
\end{equation}
where $A=(\ln\sqrt{-g})_{,0},~g=det(g_{ij}),~i,j=1,2,3$. This
force formula is not uniquely fixed rather than it depends on the
choice of frame. We define the quantity whose proper time
derivative is $F_a,~(a=0,1,2,3)$ as the momentum four-vector of
the test particle. Its spatial components give the momentum
imparted to test particles in the preferred frame (where
$g_{0i}=0$). Thus the momentum four-vector, denoted by $p_a$, is
given as [7]
\begin{equation}
p_a=\int{F_a dt},
\end{equation}
The rest of the paper is organized as follows. In the next
section, we shall describe the colliding gravitational waves in a
closed FRW background and toroidal waves. In section three, we use
the operational procedure to evaluate momentum imparted to test
particles by these waves. Finally, section four is devoted to a
brief summary of the results and concluding remarks.

\section{Gravitational Waves}
\subsection{Colliding Gravitational Waves}

The general line element describing the gravitational waves can be
written in the form [10]
\begin{equation}
ds^2=e^{-M}(dt^2-dx^2)-e^{-U}(e^{-V}dy^2+e^{V}dz^2),
\end{equation}
where $U,V$ and $M$ are functions of $t$ and $x$.

In the presence of a stiff fluid or a scalar field, the field
equations imply that $e^{-U}$ satisfies the wave equation
\begin{equation}
(e^{-U})_{tt}-(e^{-U})_{xx}=0.
\end{equation}
It is well known that the fluid can be defined in terms of a
potential function $\sigma(t,x)$ which satisfies the wave equation
\begin{equation}
(e^{-U}\sigma_t)_t-(e^{-U}\sigma_x)_x=0.
\end{equation}
In terms of this potential, the fluid density is given by
\begin{equation}
32\pi\rho=e^M(\sigma^2_t-\sigma^2_x)
\end{equation}
and its 4-velocity is proportional to the gradient of $\sigma$.

The main gravitational field equation takes the form
\begin{equation}
(e^{-U}V_t)_t-(e^{-U}V_x)_x=0
\end{equation}
and the remaining equations, which are now automatically
integrable, become
\begin{eqnarray}
U_tM_t+U_xM_x+U_{tt}+U_{xx}=\frac{1}{2}(U^2_t
+U^2_x+V^2_t+V^2_x+\sigma^2_t+\sigma^2_x),\nonumber\\
U_xM_t+U_tM_x+2U_{tx}=U_tU_x+V_tV_x+\sigma_t\sigma_x.
\end{eqnarray}
In principle, these equations can always be integrated for $M$.
Techniques for obtaining exact solutions of these equations have
been given by Carmeli et al [11].

It may be noted that the solution describing the closed FRW stiff
fluid model can be given by
\begin{eqnarray}
e^{-U}=\sin 2t\sin 2x=\sin^2(t+x)-\sin^2(t-x),~~V=\ln\tan
x,\nonumber\\ M=-\ln\sin 2t-\ln\gamma,~~ \sigma=\sqrt{3}\ln\tan t,
\end{eqnarray}
where $0<t<\pi/2$ and $0<x<\pi/2$. In this spacetime, we can
define two null hypersurfaces $t-x+a=0$ and $t+x-b=0$, where
$0<a<\pi/4$ and $\pi/4<b<\pi/2$. These can be taken to be
wavefronts of approaching gravitational waves which then collide
when $t=\frac{1}{2}(b-a)$ and $x=\frac{1}{2}(a+b)$. The FRW
solution (9) is taken only as the background region into which
these waves propagate.

\subsection{Toroidal Gravitational Waves}

The general line element of the form given by Eq.(3) can be
written in cylindrical coordinates as
\begin{equation}
ds^2=e^{-M}(dt^2-d\rho^2)-e^{-U}(e^{-V}d\phi^2+e^{V}dz^2),
\end{equation}
where $U,V$ and $M$ are functions of $t$ and $\rho$. As a result,
the vacuum field equations imply that $e^{-U}$ satisfies the wave
equation
\begin{equation}
(e^{-U})_{tt}-(e^{-U})_{\rho\rho}=0.
\end{equation}
The function $V$ satisfies the linear equation
\begin{equation}
V_{tt}-U_tV_t-V_{\rho\rho}+U_\rho V_\rho.
\end{equation}
and the remaining equations for $M$ are
\begin{eqnarray}
U_{tt}-U_{\rho\rho}=\frac{1}{2}(U^2_t
+U^2_\rho+V^2_t+V^2_\rho)-U_tM_t-U_\rho M_\rho,\nonumber\\
2U_{t\rho}=U_tU_\rho-U_tM_\rho-U_\rho M_t+V_tV_\rho.
\end{eqnarray}
If Eqs.(11) and (12) are satisfied then Eq.(13) can always be
integrated for $M$.

We can now consider a gravitational wave with toroidal wavefront
by putting [12]
\begin{equation}
U=-\ln t-\ln\rho,~~V=\ln t-\ln\rho+\widetilde{V}(t,\rho),
\end{equation}
where $\widetilde{V}$ takes the form
\begin{equation}
\widetilde{V}(t,\rho)=\int^\infty_{1/2}
{\phi(k)(t\rho)^kH_k(\frac{t^2+\rho^2-a^2}{2t\rho})}dk
\end{equation}
with an arbitrary function $\phi(k)$. Clearly these solutions
reduce to the spherical-fronted waves of [13] when $a=0$.

\section{Momentum Imparted to Test Particles}

In this section, we use the procedure outlined in the first
section to calculate momentum imparted to test particles by
colliding gravitational waves in the closed FRW spacetime
background and also the gravitational waves with toroidal
wavefronts. Using the four-vector force formula, given by Eq.(1)
in Eq.(3), we have
\begin{equation}
F_0=m[\dot{U}+\frac{\ddot{M}+2\ddot{U}}{\dot{M}+2\dot{U}}
-\frac{3\dot{U}^2+\dot{V}^2}{\dot{M}+2\dot{U}}],
~~F_1=-\frac{m}{2}M',~~F_2=0=F_3.
\end{equation}
The corresponding four-vector momentum takes the form
\begin{equation}
p_0=m[U+\ln({\dot{M}+2\dot{U}})-\int{\frac{3\dot{U}^2
+\dot{V}^2}{\dot{M}+2\dot{U}}}dt]+f_1(x),
\end{equation}
\begin{equation}
p_1=-\frac{m}{2}\int{M'}dt+f_2(x),~~p_2=constant=p_3,
\end{equation}
where $f_1(x)$ and $f_2(x)$ are arbitrary integration functions of
coordinate $x$. Here dot and prime indicate derivatives with
respect to time $t$ and coordinate $x$ respectively. It is to be
noticed that Eqs.(17) and (18) give the general result of momentum
four-vector for colliding and toroidal gravitational waves. Since
we are interested to find out the momentum imparted by these
gravitational waves, we would use Eq.(18) to calculate the term
$p_1$. As, it is obvious from this expression, we need to have the
value of $M$.

\subsection{Momentum Imparted by Colliding Gravitational Waves}

The solution describing the closed FRW stiff fluid model is given
by Eq.(9). Substituting these values in Eqs.(17) and (18), we
obtain
\begin{equation}
p_0=-m\ln(-6\cot 2t\sin 2x)+f_1(x),~~p_i=constant.
\end{equation}
We see that the momentum imparted to test particles becomes
constant which can be made zero for a special choice of constant
to be zero.

\subsection{Momentum Imparted by Toroidal Gravitational Waves}

The solution in the wave region ($t\geq \rho-a)$ can be found by
solving Eqs.(11) and (12) and is given in the form [12]
\begin{equation}
U=-\ln t-\ln\rho,~~V=\ln t-\ln \rho+\widetilde{V}(t,\rho),
\end{equation}
Using Eq.(20), we can write the equations for $M$, for $t\geq
\rho-a$, in the form
\begin{eqnarray}
M_t+M_\rho=(\frac{t-\rho}{t+\rho})(\widetilde{V_t}+\widetilde{V_\rho})
-\frac{t\rho}{2(t+\rho)}(\widetilde{V_t}+\widetilde{V_\rho})^2,\nonumber\\
M_t-M_\rho=(\frac{t+\rho}{t-\rho})(\widetilde{V_t}-\widetilde{V_\rho})
+\frac{t\rho}{2(t-\rho)}(\widetilde{V_t}-\widetilde{V_\rho})^2.
\end{eqnarray}
We now consider the case of a single component for which
$\widetilde{V}(t,\rho)$ becomes
\begin{equation}
\widetilde{V}(t,\rho)=a_k(t\rho)^kH_k(\frac{t^2+\rho^2-a^2}{2t\rho}),
\end{equation}
where $k$ is an arbitrary real parameter such that $k\geq
\frac{1}{2}$, $a_k$ is constant and $t^2+\rho^2-a^2=2t\rho$. The
parameter $k$ can also be complex, but in this case one should
consider for $\widetilde{V}$ the real or imaginary part of the
right hand side of Eq.(22). By substituting Eq.(22) into the
transformed form of Eq.(12), it is found that the functions
$H_k(\frac{t^2+\rho^2-a^2}{2t\rho})$ must satisfy the linear
ordinary differential equation which can be reduced to a
hypergeometric equation. The condition that $\widetilde{V}=0$ can
be expressed by the constraint $H_k(1)=0$. It can be seen that the
outgoing toroidal wave includes an impulsive component if the
lowest term in the expansion for $V$ has $k=\frac{1}{2}$. The
front of gravitational wave has a step (or shock) if the lowest
term has $k=\frac{3}{2}$. The value of $M$ for the wave region
$t\geq \rho-a$ turns out to be
\begin{equation}
M=\frac{1}{2k}a_k(t^2-\rho^2)(t\rho)^{k-1}H_{k-1}
-\frac{1}{2k}(t\rho)^{2k}a^2_k[k^2H^2_k
-\frac{(t^2-\rho^2)^2}{4t^2\rho^2}H^2_{k-1}].
\end{equation}
It is mentioned here that the dimension of $a_k$ is $L^{-2k}$. To
simplify the problem, we choose a particular value of $k$, i.e.,
$k=1$ for which the above equation reduces to
\begin{equation}
M=\frac{1}{2}a_1(t^2-\rho^2)H_0 -\frac{1}{2}(t\rho)^2a^2_1[H^2_1
-\frac{(t^2-\rho^2)^2}{4t^2\rho^2}H^2_0],
\end{equation}
where
\begin{eqnarray}
H_0=\ln(\frac{t}{\rho-a}),~H_1=\frac{1}{2}[(\frac{t}{\rho-a}
+\frac{\rho-a}{t})ln(\frac{t}{\rho-a})-(\frac{t}{\rho-a}
-\frac{\rho-a}{t})]\nonumber.
\end{eqnarray}
Differentiating this value of $M$ with respect to $\rho$ and
substituting in Eq.(18), then after some algebra, we arrive at the
following
\begin{eqnarray}
p_1=m\frac{a_1}{4}[\frac{t^3}{3(\rho-a)}-(\frac{\rho}{\rho-a}
+2)t\rho+2t\rho\ln(\frac{t}{\rho-a})]
+m\frac{a^2_1}{4}[\frac{1}{50(\rho-a)}(\frac{13\rho}{\rho-a}\nonumber\\
-\frac{7\rho^2}{3(\rho-a)^2}-1)t^5
+\frac{2}{9}\rho(\frac{\rho}{\rho-a}
-\frac{5}{6})t^3+\frac{1}{2}\rho(\rho(\rho-1)-(\rho-a)^2
-\frac{\rho^3}{\rho-a}\nonumber\\-4\rho^2)t
-\{\frac{1}{5(\rho-a)}(\frac{11\rho}{5(\rho-a)}
-\frac{7\rho^2}{10(\rho-a)^2}-\frac{1}{2})t^5
+\frac{2}{3}\rho(\frac{\rho}{\rho-a}
+\frac{2}{3})t^3\nonumber\\+\frac{1}{2}\rho^2(\rho-a-\frac{\rho^2}{\rho-a}
-4\rho)t\}ln(\frac{t}{\rho-a})
-\{\frac{\rho}{10(\rho-a)^2}(\frac{\rho}{\rho-a}-1)t^5
-\frac{2}{3}t^3\rho\nonumber\\-\frac{1}{2}\rho(\rho(\rho-a)+(\rho-a)^2
-2\rho^3)t\}(ln(\frac{t}{\rho-a}))^2+f_2(\rho).
\end{eqnarray}
This is the momentum imparted to test particles by gravitational
waves with toroidal wavefronts. We see that the momentum term
approaches to zero in the limit $t\rightarrow 0$ for the
particular choice of $f_2=0$. This result reduces to the momentum
imparted by gravitational waves with spherical wavefronts for
$a=0$ [8]. The interpretation of the term $p_0$ of the four-vector
momentum has been discussed in detail elsewhere [14].

The character of the gravitational wave near the wavefront can be
determined by the component of $\widetilde{V}$ with the minimum
value of $k$. It can be shown that, near the wavefront as
$t\rightarrow\rho-a$,
\begin{equation}
\widetilde{V}\sim a_k(t+\rho)^{-1}(t-\rho+a)^{1+2k},
~M\sim((1+2k)/2k)a_k(t-\rho+a)^{2k}.
\end{equation}
Making use of this value of $M$ in Eq.(18), we obtain
\begin{equation}
p_1=\frac{m}{4k}(1+2k)a_k(t-\rho+a)^{2k}+f_2(\rho).
\end{equation}
This is the momentum imparted to test particles near the toroidal
wavefront.

\section{Discussion}

We have evaluated momentum imparted to test particles by colliding
gravitational waves in a closed FRW background and also for
toroidal gravitational waves by using an entirely different
approach. It can be seen that the momentum becomes constant for
colliding gravitational waves in the closed FRW background. This
is what one would expect from this procedure. It also follows from
Eq.(25) that the momentum imparted by toroidal gravitational waves
is physically well behaved. This turns out to be zero in the limit
$t\rightarrow 0$ for a particular value of arbitrary function. We
also see that this becomes singular at $\rho=a$. If we take $a=0$,
it reduces to the result of spherical wavefronts [8] and the
singularity shifts to $\rho=0$ which also acts as a source of the
gravitational waves inside the wave region. It is interesting to
note that all these results exactly coincide with the results
obtained by using M$\ddot{o}$ller prescription [15] for a
particular choice of arbitrary function $f_2$.

\newpage

\begin{description}
\item  {\bf Acknowledgment}
\end{description}

The author would like to thank Ministry of Science and Technology
for postdoctoral fellowship at University of Aberdeen, Scotland,
UK.

\vspace{2cm}

{\bf \large References}

\begin{description}

\item{[1]} Cooperstock, F.I.: Found. Phys. {\bf 22}(1992)1011; in
{\it Topics in Quantum Gravity and Beyond: Papers in Honor of L.
Witten} eds. Mansouri, F. and Scanio, J.J. (World Scientific,
Singapore, 1993) 201; in {\it Relativistic Astrophysics and
Cosmology}, eds. Buitrago et al (World Scientific, Singapore,
1997)61; Annals Phys. {\bf 282}(2000)115.

\item{[2]} Misner, C.W., Thorne, K.S. and Wheeler, J.A.: {\it
Gravitation} (W.H. Freeman, San Francisco, 1973).

\item{[3]} Ehlers, J. and Kundt, W.: {\it Gravitation: An
Introduction to Current Research}, ed. L. Witten (Wiley, New York,
1962)49.

\item{[4]} Pirani, F.A.E.: {\it Gravitation: An Introduction to
Current Research}, ed. L. Witten (Wiley, New York, 1962)199.

\item{[5]} Weber, J. and Wheeler, J.A.: Rev. Mod. Phy. {\bf 29}(1957)509;\\

\item{[6]} Weber, J.: {\it General Relativity and Gravitational
Waves}, (Interscience, New York, 1961).

\item{[7]} Qadir, A. and Sharif, M.: Physics Letters {\bf A
167}(1992)331.

\item{[8]} Sharif, M.: Nuovo Cimento {\bf B 116}(2001)1311.

\item{[9]} Qadir Asghar and Sharif, M.: Nuovo Cimento {\bf B
107}(1992)1071;\\
Sharif, M.: Ph.D. Thesis, Quaid-i-Azam University Islamabad
(1991).

\item{[10]} Feinstein, A. and Griffiths, J.B.: Class. Quantum
Grav. {\bf 11}(1994)L109.

\item{[11]} Carmeli, M., Charach, Ch. and Feinstein, A.: Ann.
Phys. {\bf 150}(1983)392.

\item{[12]} Alekseev, G.A. and Griffiths, J.B.: Class. Quantum
Grav. {\bf 13}(1996)2191.

\item{[13]} Alekseev, G.A. and Griffiths, J.B.: Class. Quantum
Grav. {\bf 12}(1995)L-13.

\item{[14]} Qadir, Asghar, Sharif, M. and Shoaib, M.: Nuovo
Cimento {\bf B 115}(2000)419.

\item{[15]} M$\ddot{o}$ller, C.: Ann. Phys. (NY) {\bf 4}(1958)347;
{\bf 12}(1961)118.

\end{description}

\end{document}